\documentclass[12pt]{article}
\usepackage{newtxtext,newtxmath}

\usepackage{graphicx}
\usepackage[dvipsnames]{xcolor}
\usepackage{soul}

\usepackage[letterpaper,margin=1in]{geometry}

\usepackage{siunitx}

\renewenvironment{abstract}
	{\quotation}
	{\endquotation}

\date{}

\makeatletter
\renewcommand{\fnum@figure}{\textbf{Figure \thefigure}}
\renewcommand{\fnum@table}{\textbf{Table \thetable}}
\makeatother

\usepackage{scicite}
\usepackage{url}

\definecolor{taupe}{RGB}{125,87,46}

\def\scititle{
Hydrodynamics converts chiral flagellar rotation into contactless actuation of microdiscs\\
}

\title{\bfseries \boldmath \scititle}

\author{
	Daniel Grober$^{1}$,
	Tanumoy Dhar$^{2}$,
	David Saintillan$^{2}$,
        Jeremie Palacci$^{1\ast}$,\and
	\small$^{1}$Institute for Science and Technology, Austria, Klosterneuburg, Austria\and
	\small$^{2}$ Department of Mechanical and Aerospace Engineering, University of California San Diego, CA, USA\and
	\small$^\ast$Corresponding author. Email: jeremie.palacci@ista.ac.at\and
}

\begin{document} 

\maketitle


\begin{abstract} \bfseries \boldmath

Motile bacteria are a wonder of nature’s engineering: microscopic engines that transduce biochemical energy into the work they require to explore their environment. This added energy turns the surrounding fluid into a bath that departs from an equilibrium one. Bacterial baths agitate suspended spheres more vividly than thermal fluctuations and can power microscopic ratchets. 
A salient requirement to extract work from bacterial baths was the asymmetric shape of the ratchets, designed to rectify the interactions with bacteria.
In contrast with past results, here we show that swimming {\it E. coli} power the persistent rotation of discs, in absence of asymmetry. Combining state-of-the art nanoprinting, quantitative measurements of the dynamics of individual bacteria, and hydrodynamic modeling, we elucidate the mechanism  and show that the counter-rotation of the flagella and the bacterium body lead to a torque dipole and traction onto the disc, and subsequent rotation. Remarkably, the mechanism is independent of the direction or orientation of navigation of bacteria under the disc, hence additive and contactless. 
Resulting from the interplay of the torque dipole of flagellated bacteria with simple geometric confinement, this hydrodynamic mechanism  bridges scales, leveraging the chirality of bacteria nanomotors towards the manipulation of objects at least ten thousand times larger. 
The study lays the groundwork for novel bio-hybrid micromachines that harness living microorganisms for controlled motion at the microscale. 
Our findings provide further fundamental insights into bacterial hydrodynamics and open avenues for the development of autonomous, self-powered microdiscs for the study of chiral fluids.

\end{abstract}

\noindent
\section*{Introduction}
Motile bacteria are micron-sized, self-propelling machines that convert biochemical energy from carbon and oxygen sources into mechanical work and motion. Owing to the kinematic reversibility of motion at low Reynolds numbers, the propulsion mechanism must break time-reversal symmetry to achieve net displacement \cite{Purcell.AJP.1977}.  For {\it E. coli} bacteria, this is achieved with the actuation of flagella by the proton-driven Bacteria Flagellar Motor (BFM)\cite{Manson.PNAS.1977}, whose persistent rotation propels the bacteria forward \cite{Berg.NAT.1972, Berg.PT.2000}. This results in the transfer of mechanical work from the bacteria to the fluid, taking the surrounding medium out of equilibrium, effectively an active bath. It leads, for example, to an increase in the diffusivity of colloidal spheres controlled by the activity of the bacteria \cite{Wu.PRL.2000, Leptos.PRL.2009, Mino.PRL.2010, Jepson.PRE.2013}. Remarkably, in an apparent departure from equilibrium physics, bacterial baths power microgears, provided they present asymmetric shapes \cite{DiLeonardo.PNAS.2010, Sokolov.PNAS.2010,Angelani.2008} that rectify the forces exerted by bacteria against the walls of the gears. In those experiments \cite{DiLeonardo.PNAS.2010, Sokolov.PNAS.2010}, symmetric gears simply fluctuate, while asymmetric gears display persistent rotation, whose direction is controlled by geometry \cite{Vizsnyiczai:2017ip}. This strategy was further replicated with active colloids nested in asymmetric gears \cite{Maggi:2015dx} and used to control the propulsion of objects in bacterial baths \cite{RdL_arxiv2025, Koumakis.NatComm.2013, Steager.JMM.2011, Carlsen.LabOChip.2014}.
Alternatively, we recently demonstrated that colloidal aggregates exhibited persistent rotation when immersed in a suspension of swimming {\it E. coli}. We reported that the direction of rotation was controlled by the slip conditions of the interface where aggregates sit. Colloidal aggregates rotate clockwise on the no-slip surface of a glass capillary and counter-clockwise on an air-water interface, reminiscent of the change of direction of rotation of swimming {\it E. coli} on interfaces \cite{DiLeonardo:2011cpb}. Our findings pointed towards a coupling of the rotating BFM with the aggregates as key to the rotation. It resulted that suspensions of swimming {\it E. coli} are chiral bacterial baths that transmit torques to colloidal aggregates, which ultimately controls the formation of unconventional aggregates and gels\cite{Grober.NatPhys.2023}. Because colloidal aggregates arise from collisions and irreversible bindings, they present indiscriminate shapes, thus lacking the aforementioned symmetry to achieve persistent rotation as previously reported\cite{DiLeonardo.PNAS.2010, Sokolov.PNAS.2010, Vizsnyiczai:2017ip}. 
Further investigations with model experiments are therefore required to elucidate the effect of swimming {\it E. coli} on symmetric objects. To this end, we leverage state-of-the-art 3D nanoprinting and study the effect of suspensions of motile {\it E. coli} on micrometric and symmetric discs [Fig. 1A,B]. We notably devise two sets of circular discs, dubbed ``pucks" [Fig. 1C,D]. The first set consists of simple thick discs. The second set is composed of discs, as in the first set, with the addition of a narrow channel along the diameter, that allows bacteria to cross the puck.
Here, we show that the BFMs of the swimming {\it E. coli} exert a torque dipole and a traction force that leads to the rotation of even symmetric objects.
Our experimental findings are described by a hydrodynamic model of swimming bacteria as torque dipoles that quantitatively predicts the rotation rates observed in experiments.  Remarkably, the effect is additive, with each present bacteria contributing a torque; we leverage this effect to power bio-hybrid machines and demonstrate first steps towards the realization of a chiral fluids of spinners.

\section*{Results}
\subsection*{Bacteria are torque dipoles: swimming in circles}
As our observations are performed near the bottom of a glass capillary, where objects are confined by gravity, we first recapitulate the swimming behavior of swimming {\it E. coli} near a no-slip wall. Swimming {\it E. coli} are force and torque-free. They are accurately represented hydrodynamically by a force dipole \cite{10.1073/pnas.1019079108} and a torque dipole: the flagella spinning one way and the head spinning the other way to balance the torque. In effect, swimming {\it E. coli} are hydrodynamically attracted to solid walls by the image charge of the force dipole \cite{10.1103/PhysRevLett.101.038102} and swim in (clockwise) circles as a result of the opposing shear forces induced by the torque dipole \cite{Lauga.BJ.2006} [Fig.~\ref{figSTrajectories}A].

\subsection*{Experimental procedure}
We turn to the dynamics of 3D-printed pucks in the presence of swimming bacteria. The pucks are printed with radius $R=5$, 10, \SI{20}{\micro\meter} and constant height $\sim$\SI{6}{\micro\meter} using a Two-Photon-Polymerization (2PP) printer (NanoOne, Upnano) [Fig.~\ref{fig1}A, SI]. After print and development, they are dispersed in a solution of 5$\%$ F-108 surfactant to prevent aggregation and subsequently concentrated [Fig~\ref{figSPucks}]. The pucks are added to a suspension of swimming {\it E. coli} in motility medium and sealed in a glass capillary (see Materials and Methods). The concentration of swimming {\it E. coli} ($\rho_B$) is adjusted prior to the experiment,  as described in each section. The pucks sediment, sitting at the bottom of the capillary, and interact with swimming \textit{E. coli} [Fig.~\ref{fig1}B]. We carry out observations of the pucks and bacteria by fluorescence microscopy, using the autofluorescence of the nanopriting resin and the green fluorescent protein (GFP) tagging of the bacteria [SI]. A dot and a line are added to the design to track the orientation of the discs.

\subsection*{Dynamics of a simple puck in a bacterial bath of swimming {\it E. coli}}
We first characterize the dynamics of the simple pucks: thick discs, in a bacterial bath. We use a concentration $\rho_B = 6\times 10^8$ cells/mL as in \cite{Grober.NatPhys.2023}, where clockwise rotation of aggregates was observed. Bacteria collide with the perimeter of the puck and deflect, but do not cross underneath [Movie S1, Fig.~\ref{figSTrajectories}B]. 
The pucks exhibit noisy dynamics at short times, reminiscent of the high effective temperature of the bacterial bath \cite{Wu.PRL.2000, Leptos.PRL.2009, Mino.PRL.2010, Grober.NatPhys.2023}.
Over the course of minutes, however, the pucks  display a  slow but perceptible clockwise rotation, as observed with colloidal aggregates. We repeat the experiment for pucks of radii: \SI{5}{\micro\meter}, \SI{10}{\micro\meter}, and \SI{20}{\micro\meter} (at constant thickness $\sim$ \SI{6}{\micro\meter}) and observe clockwise rotation at all sizes [Fig.~\ref{fig1}E]. We quantify our observations by tracking the angle $\Theta (t)$ [Fig.~1B] and compute the rotation rate $\omega_R$ of the puck by linear fit of $\Theta (t)$. We notice that small pucks spin faster than larger ones following a scaling $\omega_R\propto 1/R$ [Fig.~\ref{fig1}F].
We next quantify the noise acting on the pucks, observing diffusive dynamics at short times [Fig.~\ref{figSMSAD}] with rotational diffusivity $D_{\Theta}\propto 1/R^3$
[Fig.~\ref{fig1}F -inset]. The observed scalings for the rotational diffusivity, $D_{\Theta}\propto 1/R^3$ and rotation rate of the pucks $\omega_R\propto 1/R$ are in line with our previous measurements with colloidal aggregates \cite{Grober.NatPhys.2023}, indicating an active torque generated by the bacterial bath, $\Gamma \propto R^2$. Its physical origin can be intuited with a toy model, accounting for collisions of bacteria with chiral and curved trajectories onto the perimeter of a disc [Fig.~\ref{figSTrajectories}]. In brief, the curvature of the clockwise trajectories leads to asymmetric collisions, and to a net torque $\Gamma \propto R^2$ resulting in clockwise rotation [see \cite{Grober.NatPhys.2023} for details on this toy model]. In effect, this experiment shows that collisions of bacterial baths of swimming {\it E. coli}, exemplified on [Fig.~\ref{figSTrajectories}], can produce active torques on symmetric objects by interacting with their perimeter.

\subsection*{Dynamics of a puck with a channel }
\subsubsection*{Comparison of the dynamics of a puck with and without a channel }
We now investigate the effect of {\it E. coli} swimming under objects, using pucks with a square channel, \SI{2}{\micro\meter} x \SI{2}{\micro\meter},  crossing the puck along a diameter  [Fig.~\ref{fig1}D]. We only consider situations where the channel lays on the bottom substrate, facing down. Swimming bacteria enter the channel and proceed to the exit of the puck, the tight confinement preventing them from reversing course. 
In Fig.~1G, we compare the dynamics of 2 different puck designs (both are \SI{6}{\micro\meter} thick disks, one without and one with a channel), immersed at once in the same bath of motile \textit{E. coli} at concentration $\rho_B = 6\times 10^8$ cells/mL; the blue line represents a puck without a channel, while the black line represents a puck with a channel.
For both pucks, we observe a persistent clockwise rotation. 
However, pucks with a channel display more complex dynamics. Without \textit{E. coli} in the channel, they slowly spin clockwise, at rates comparable to the pucks without a channel; their rotation arises solely from collisions with the perimeter [Fig.~1G, black and blue lines]. 
The rotation rate momentarily increases as a single \textit{E. coli} crosses the channel [Fig.~1G, green dots].
However, we observe four events of marked, long increase in the rotation rate [Fig.~1G, red dots]. They correlate with events of multiple \textit{E. coli} entering the channel from opposite sides that become jammed and halted. 
These observations contrast with previous reports of rotation of asymmetric gears, ruling out collisions of bacteria with walls as the driving mechanism \cite{DiLeonardo.PNAS.2010, Sokolov.PNAS.2010}.

\subsubsection*{Dynamics of a puck with a channel crossed by a single swimming {\it \textbf{E. coli}}}
To elucidate the origin of this rotation and the effect of the swimming bacteria inside the channel, we consider the dynamics of a puck while its  channel is crossed by a {\it single} swimming {\it E. coli}. To this end, we suspend the  pucks (with channels) in a dilute bacterial bath, where the density $\rho_B = 3\times 10^7$ cells/mL is 20 times lower than previously. At this concentration, we can investigate the effect of a single bacterium crossing through the channel, in the absence of collisions of swimming bacteria with the perimeter of the puck [Movie S2]. Timelapse acquisition is performed by spinning disk confocal fluorescence microscopy (Nikon TI-2 Eclipse, 10 fps, See Materials and Methods) and analyzed to record simultaneously the position of the center of mass of the puck, the orientation $\Theta$ of the channel and the position of the body of the bacterium crossing the channel. It is important to note here that only the body of the bacteria is fluorescently labeled, and that the flagella, a floppy tail of $\sim~$\SI{6.5}{\micro\meter} \cite{10.1128/jb.06735-11}, cannot be observed in our experiments.  We quantify our experimental observations by representing the change of angle $\Delta \Theta$ of the puck after entry of the bacterium in the channel, as a function of the position $X_B$ of the center of mass of the body of the bacterium in the channel.  This representation allows us to collapse data from bacteria with different swimming velocity [Fig.~\ref{fig2}A-D] as expected from low-Reynolds-number dynamics. Indeed,  reversibility of Stokes flow \cite{Purcell.AJP.1977} dictates that the net motion of the puck is independent of the rate, {\it i.e. velocity}, at which bacteria cross the channel. In effect, while bacteria with different swimming speeds [Fig.~2A] present different timelapse of crossing [Fig.~2B], the dynamics of the puck is collapsed when represented as a function of the position of the bacterium in the channel, $\Delta\Theta (X_B)$ [Fig.~\ref{fig2}C].
Initially, the puck rotates clockwise, and reverses direction when the \textit{E. coli} is roughly one body length away from the exit of the channel [Fig.~\ref{fig2}C, black dashed line]; we refer to this up-down shape as a ``swoosh". 
Remarkably, the characteristic shape of the swoosh is unchanged whether the bacterium enters the channel from one end or the other. While the $\Delta\Theta (X_B)$ representation effectively collapses the dynamics of rotation of the pucks for similarly sized bacteria with different swimming velocity [Fig.~\ref{fig2}A-C], it shows noticeable differences in the depth ($\Delta \Theta_{max}$) and position of the swoosh for bacteria of different body length ($\ell_B$) [Fig.~2G]. Those differences do not correlate with the average angle of the cell body with respect to the channel [Fig.~\ref{figSCorr}].  
Instead, the depth of the swoosh ($\Delta \Theta_{max}$) correlates with the size of the bacterium body ($\ell_B$) [Fig.~\ref{fig2}E,F], quantified by fluorescence imaging of the body. When bacteria with a longer body cross the channel, the swoosh is deeper ({\it i.e.} larger $\Delta \Theta_{max}$), and the reversal of direction occurs earlier ({\it i.e.} $X_B$ further from the exit).

\subsection*{Hydrodynamic model}
The aforementioned observations rule out collisions of the bacteria with the inner channel walls---the driving force behind the rotation of asymmetric gears \cite{DiLeonardo.PNAS.2010, Sokolov.PNAS.2010}---as a potential mechanism for the rotation of the pucks with a channel. Instead, we recall that swimming \textit{E. coli} cells exert a torque dipole on their surrounding, stemming from the counter rotation of the cell head (clockwise when viewed from the rear) and flagella (counter-clockwise), and we intuit that these applied torques lead to the observed phenomenology. The rotation of the head entrains the fluid around it, resulting in a traction field (shear stress) on the walls of the channel. The counter-rotation of the flagella similarly generates an oppositely directed traction field. Because they oppose each other, the two traction fields do not result in a net force on the puck; however, since they are displaced along the channel axis by the effective length of the torque dipole $\ell_D$---a distance on the order of the bacterial length, they can apply a net torque on the puck and drive its rotation [Fig.~3A].

To confirm this mechanism, we model the hydrodynamic interaction of a single \textit{E. coli} cell swimming through a square microchannel of width $2W$.  The bacterium is assumed to be aligned with the axis of the channel, consistent with experimental observations. For analytical progress, we approximate the channel walls as infinite stationary boundaries. To leading approximation, the bacterium exerts both a force dipole and a torque dipole on the fluid around it. For a bacterium aligned with the channel axis ($x$ direction), symmetry precludes the force dipole from driving any net torque, and we therefore omit it in our flow calculation. Instead, we idealize the bacterium as exerting two equal and opposite point torques (or rotlets) $\pm \Gamma_M\, \hat{\mathbf{x}}$ at locations $\mathbf{r}_{1,2}$ offset by a fixed distance $\ell_D$ along the channel axis: $\mathbf{r}_1-\mathbf{r}_2=\ell_D\, \hat{\mathbf{x}}$. The microscopic torque magnitude $\Gamma_M$ is given by the bacterial motor torque, while the torque spacing, or dipole length $\ell_D$, is expected to scale with the size of the bacterium, a point we elaborate on further below. 

We first analyze the effect of a single rotlet $+\Gamma_M \hat{\mathbf{x}}$ at location $\mathbf{r}_1$. At low Reynolds number, the fluid motion it induces inside the channel satisfies the Stokes equations,
\begin{equation}\label{eq:stokes}
    \mathbf{\nabla \cdot U} = 0, \, \quad \mathbf{\nabla \cdot \boldsymbol{\Sigma}} = -\frac{\Gamma_M}{2}\nabla \times [\delta(\mathbf{r}-\mathbf{r}_1)\hat{\mathbf{x}}]  ,   
\end{equation}
where $\mathbf{U}$ is the fluid velocity, $\boldsymbol{\Sigma} = -P\mathbf{I} + 2\mu \mathbf{E}$ is the Newtonian stress tensor expressed in terms of the pressure $P$, dynamic viscosity $\mu$, and rate-of-strain tensor ${\mathbf{E}}=\tfrac{1}{2}(\nabla {\mathbf{U}}+ \nabla{\mathbf{U}}^{\mathrm{T}})$, and $\delta(\mathbf{r})$ is the Dirac delta function. The fluid velocity is subject to the no-slip condition at the channel walls: $\mathbf{U}(x, y=\pm W, z=\pm W) = \mathbf{0}$, where the $x$ coordinate is aligned with the channel axis and the $z$ direction points normal to the bottom substrate [Fig.~3A]. By linearity of the Stokes equations, the fluid velocity depends linearly on the torque, 
\begin{equation}
\mathbf{U}(\mathbf{r})= \frac{1}{8\pi\mu} \mathbf{R}(\mathbf{r}-\mathbf{r}_1)\cdot \Gamma_M \hat{\mathbf{x}},   \label{eq:rotletvel}
\end{equation}
where $\mathbf{R}(\mathbf{r})$ is the Green's function for this problem. As explained in [SI], the solution for $\mathbf{U}(\mathbf{r})$ can be obtained numerically by solving equation Eq.~(\ref{eq:stokes}) using the boundary-element method \cite{pozrikidis1992boundary} [Fig. 3B]. The velocity field Eq.~(\ref{eq:rotletvel}) exerts a traction on the channel walls. Since the bottom wall belongs to the fixed substrate, only viscous stresses on the top wall ($z=+W$) contribute to the vertical torque on the puck. There, the viscous traction is
\begin{equation}
\mathbf{t}(\mathbf{r})= -\hat{\mathbf{z}}\cdot 2\mu \mathbf{E}(\mathbf{r})= -\mu \left(\frac{\partial U_x}{\partial z}\hat{\mathbf{x}}+\frac{\partial U_y}{\partial z}\hat{\mathbf{y}}   \right),  \qquad z=+W.   \label{eq:traction}
\end{equation}
It results in a net torque on the puck given by 
\begin{equation}
\Gamma_1\hat{\mathbf{z}}= \int_{z=+W} (\mathbf{r}-\mathbf{r}_C)\times \mathbf{t}(\mathbf{r}-\mathbf{r}_1)\,\mathrm{d} S = -(x_1-x_C) \int_{z=+W} \mu\frac{\partial U_y}{\partial z}\mathrm{d}S\,\hat{\mathbf{z}}\,,
\end{equation}
where $\mathbf{r}_C$ denotes the center of the puck. 
Upon inserting Eq.~(\ref{eq:rotletvel}), the torque magnitude reduces to
\begin{equation}  \label{eq:gamma1}
\Gamma_1 = -\Lambda \left(\frac{x_1-x_C}{W}\right) \Gamma_M\,, \quad \mathrm{where}\quad  \Lambda = \frac{W}{8\pi}\int_{z=+W} \frac{\partial R_{yx}}{\partial z}\mathrm{d}S.
\end{equation}
This expression captures the viscous torque transmission from the point rotlet ($\Gamma_M$) to the puck ($\Gamma_1$). Note that $\Lambda$ is a positive dimensionless constant independent of any parameters (including $W$); our boundary-element calculations in an infinite square channel provide a value of $\Lambda\approx 0.17$.

As the bacterium swims through the channel, the torque dipole resulting from the counter-rotation of the cell body and flagella yields a net torque on the puck given by  
\begin{equation}
\Gamma\hat{\mathbf{z}}=\Gamma_1\hat{\mathbf{z}}+\Gamma_2\hat{\mathbf{z}}=-\Lambda \frac{\ell_D}{W}\Gamma_M \hat{\mathbf{z}}\,,  \label{eq:nettorque}
\end{equation}
where $\ell_D=x_1-x_2$ is the dipole length. Remarkably, the torque magnitude $\Gamma$ is independent of the position of the bacterium under the puck, provided that both rotlets are located inside the channel; it is also independent of the orientation of the bacterium ($+\hat{\mathbf{x}}$ or $-\hat{\mathbf{x}}$) along the channel axis, as observed in the experiment.

We can now describe the angular dynamics of the puck. We denote by $X_B(t)=U_s t$ the instantaneous position of the bacterium inside the channel, measured from the channel entrance. Here, $U_s$, the bacterial swim speed, is constant as measured experimentally [Fig.~2A]. So long as both rotlets are contained inside the channel, the torque on the puck is constant and given by equation Eq.~(\ref{eq:nettorque}), resulting in the angular velocity
\begin{equation}
\frac{\mathrm{d}\Theta}{\mathrm{d}t}= {M}_\Theta \Gamma,  \label{eq:angvel}
 \end{equation}
where ${M}_\Theta$ is the rotational mobility of the puck for rotations around the $z$ axis. The value of ${M}_\Theta$ is obtained from the Stokes-Einstein relation, ${M}_\Theta=D_\Theta /k_B T$, where the rotational diffusivity of the puck in a thermal bath was measured independently: $D_{\Theta} = (6\pm 1) \times10^{-5}\,\, \text{rad}^{2}/\text{sec}$ [Fig.~\ref{figSPuckThermal}]. Integrating Eq.~(\ref{eq:angvel}) and eliminating time using the swim speed provides the angular displacement as a function of the position $X_B$ of the bacterium in the channel:
\begin{equation}
\Delta \Theta(X_B) = -\Lambda \frac{\ell_D}{W}\frac{{M}_\Theta}{U_s} \Gamma_M X_B \,.   \label{eq:Theta}
\end{equation}
This relation predicts clockwise rotation of the puck and captures the linear decrease observed in the experimental data of [Fig.~2C,G]. All the prefactors in Eq.~(\ref{eq:Theta}) can be estimated based on experiments [Table S1], with the exception of the dipole length $\ell_D$. The collapse of the angular displacements upon scaling $\Delta \Theta$ with cell body length in [Fig.~2F] points at a linear relationship between $\ell_D$ and $\ell_B$, and therefore we posit that $\ell_D=\alpha \ell_B$. The dimensionless parameter $\alpha$ is the only fitting parameter of our model. By fitting Eq.~(\ref{eq:Theta}) to the experimental data, we estimate $\alpha \approx 1.5$.

This simple hydrodynamic model also allows us to explain the counter-clockwise swoosh observed in [Fig.~2C,G]. As the cell body exits the channel, it ceases to exert a torque on the puck, which is now only subject to the torque $\Gamma_2$ due to the rotating flagella, thus causing a change in the direction of rotation. We estimate the angular displacement beyond that point to be 
\begin{equation}
\Delta \Theta(X_B) = \Lambda \frac{1}{W}\frac{{M}_\Theta}{U_s} \Gamma_M  \left[\frac{X_B^2}{2}+X_B(\ell_B -\ell_{D}-R) + \frac{\ell_B(\ell_B-2R)}{2} \right],
\end{equation} 
which predicts a reversal in the direction of rotation with a quadratic dependence on position. 
The hydrodynamic model provides a quantitative description of the rotation through  Eq. (8) and (9), controlled by the presence of either the two rotlets or a single one inside the channel. The transition is observed in the experiment as the position of reversal of the swoosh [Fig.~\ref{fig2}C].

\subsection*{Discussion}
We now inspect the physical implications of this hydrodynamic mechanism. 
First, the entrainment of the puck by a torque dipole explains the swoosh shape observed in the experiment: when both torque rotlets are present in the channel, the puck rotates clockwise, with a direction of rotation that reverses when a single rotlet remains in the channel. In effect, the model quantitatively reproduces our experimental findings [Fig.~3C], capturing both the reversal of the direction of rotation and the depths $\Delta \Theta _{max}$ observed in the experiment.  In the experiment, the swoosh, {\it i.e.} reversal of the direction of rotation of the puck, is observed near $X_B\sim 2R-\ell_B$ [Fig.~2D]. 
Naively, we expected the reversal of direction of the swoosh to occur at $X_B = 2R$, when the center of the body of \textit{E. coli} exits the channel.
The discrepancy suggests that our description, notably assuming infinite channels and a far-field rotlet hydrodynamic description, will require additional refinements to better characterize the dynamics of the bacterium exiting the channel. Finally, the experimental observations confirm our intuition that the dipole length is linearly dependent on the size of the bacterium body, as all experiments can be fitted with a single parameter $\alpha$ such that $\ell_D=\alpha\ell_B$.
Remarkably, while the bacteria exert a torque on the puck throughout the navigation, the recovery stroke nearly cancels out the initial rotation after a complete crossing event. We realize that persistent rotation would be achieved by maintaining both rotlet dipoles in the channel, {\it i.e.} by preventing the bacterium body from exiting the channel. To investigate this hypothesis, we print pucks
with closed channels that terminate at short distance from the center [Fig.~4A, Movie S3]. This geometry aims to trap bacteria in full (body and flagella) in the closed channel and to suppress additional torques that arise from \textit{E. coli} pushing on the end wall of the channel. As hypothesized, those pucks start rotating clockwise as soon as bacteria are trapped inside one of the channels. Pucks rotate faster as their occupancy by bacteria increases, irrespective of the relative position of the bacteria in the channels [Fig.~4A,C],  highlighting the additive nature of our contactless rotation mechanism. The clockwise rotation persists of over the course of minutes, at speeds that compare with asymmetric ratchets in bacterial baths [Fig.~4A,B]\cite{DiLeonardo.PNAS.2010, Sokolov.PNAS.2010}.  Bacteria with both body and flagella in a channel contribute each a net torque of the same sign that cumulatively rotate the puck in the same (clockwise) direction [Fig.~4D]. Once again, this contrasts with mechanisms driven by bacteria pushing on asymmetric gears, where the geometric positioning of the bacteria was essential. Guided by our hydrodynamic model, we realize that the total length of trapped bacteria rather than their number should set the rotation and we collapse the observed rotation rates of the puck $\omega\propto \Sigma_i \ell^i_B$, where $\ell^i_B$ is the measured length of the body of the bacterium $i$ in the channel [Fig.~4B]. 
Furthermore, we show that our hydrodynamic model accurately predicts the observed rotation speeds by simply modifying the hydrodynamics of an open channel onto a closed one [Fig.~4D].

To conclude, we unveiled a hydrodynamic, contactless  mechanism that leads swimming {\it E. coli} to rotate symmetric objects. It originates from the chirality of the flagella nanomotors, bridging orders of magnitude of scales through hydrodynamics and confinement. Our findings open new opportunities to design hybrid, \textit{E. coli} - powered, microdiscs that persistently rotate in the absence of geometric asymmetry. The rotation can be tuned by modifying the number of channels, and scaled up by using commercially available 2PP nanoprinters, making it possible to study, for example, chiral fluids of spinners, powered from within. As a first step in this direction, we show a collection of discs powered by a bacterial bath [Fig. 4E, Movie S4]. Our experimental results notably highlight that the capture of multiple bacteria in narrow channels [Fig. 4F-inset] increase the rotation rate up to 10 RPM [Fig.~4F], pointing out the salient role of confinement to enhance the hydrodynamic effect. Our findings highlight the interplay between bacterial navigation and their environments, with potential ecological implications and a novel route to assemble unconventional gels in bacterial baths \cite{Grober.NatPhys.2023}.


\newpage


\begin{figure} 
	\centering
	\includegraphics[width=0.85\textwidth]{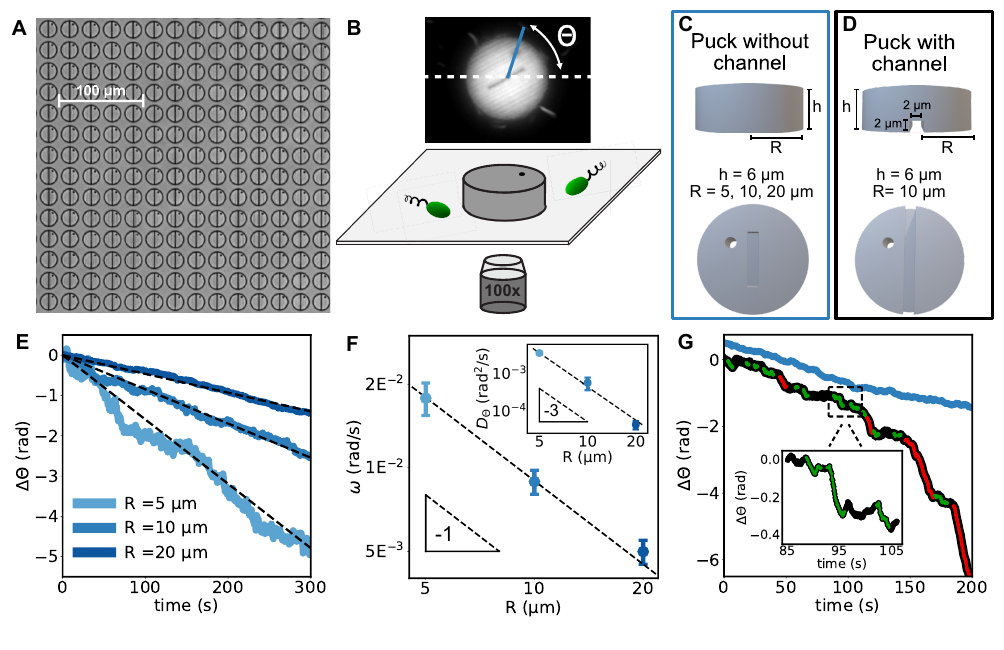} %

	\caption{\small \textbf{Dynamics of colloidal pucks in a bath of swimming \textit{E. coli}} ($\rho_B = 6\times 10^8$ cells/mL).
		(\textbf{A}) Micron-scale discs, dubbed pucks, are printed in 50x50 arrays onto a glass cover slip using a 3D nanoprinter and subsequently suspended.
            (\textbf{B}) Schematic representation of the experiment, where pucks are immersed in a bath of swimming \textit{E. coli} bacteria. (Top picture) Fluorescent microscopy image of a puck and definition of its orientation $\Theta$.
            (\textbf{C-D}) Designs of the  pucks. They show a dot as marker to track the rotation the pucks. (\textbf{C}) Simple puck:  disc  of height \SI{6}{\micro\meter} and radii = 5, 10, and \SI{20}{\micro\meter}. 
            (\textbf{D}) Puck with a channel: discs of  height \SI{6}{\micro\meter} with an added square (\SI{2}{\micro\meter} x \SI{2}{\micro\meter}) crossing channel. 
            (\textbf{E}) Dynamics  of pucks,  $\Theta(t)$, for radii : \SI{5}{\micro\meter} (light blue), \SI{10}{\micro\meter} (blue), and \SI{20}{\micro\meter} (dark blue).   All pucks exhibit a slow clockwise rotation, whose linear fit measures the average rotation rate $\omega$ (dashed black lines). 
            (\textbf{F}) Puck rotation rate $\omega$; error-bars represent repeated experiments using different pucks of the same radius. 
            (\textbf{F-inset}) Rotational diffusion for pucks of difference sizes, measured from mean squared angular displacement [SI];
            error-bars represent repeated experiments using different pucks of the same radius. 
            (\textbf{G}) Comparison of the dynamics for a simple puck, (panel C) in blue (offset for readability) or a puck with a channel (panel D) in black, in a bacterial bath. Both pucks rotate clockwise. For pucks with a channel, single bacteria crossing the channel (green dots) momentarily increase the rotation rate (inset). We observe 4 events of marked, long increase of rotation rates (red dots) that correlate with multiple bacteria entering from opposite sides of the channels, jamming and halting inside the channel [see main text]. Those observations rule out collisions of bacteria with the inner wall as the origin for the rotation.
            }
	\label{fig1} 
\end{figure}

\begin{figure} 
	\centering
	\includegraphics[width=0.62\textwidth]{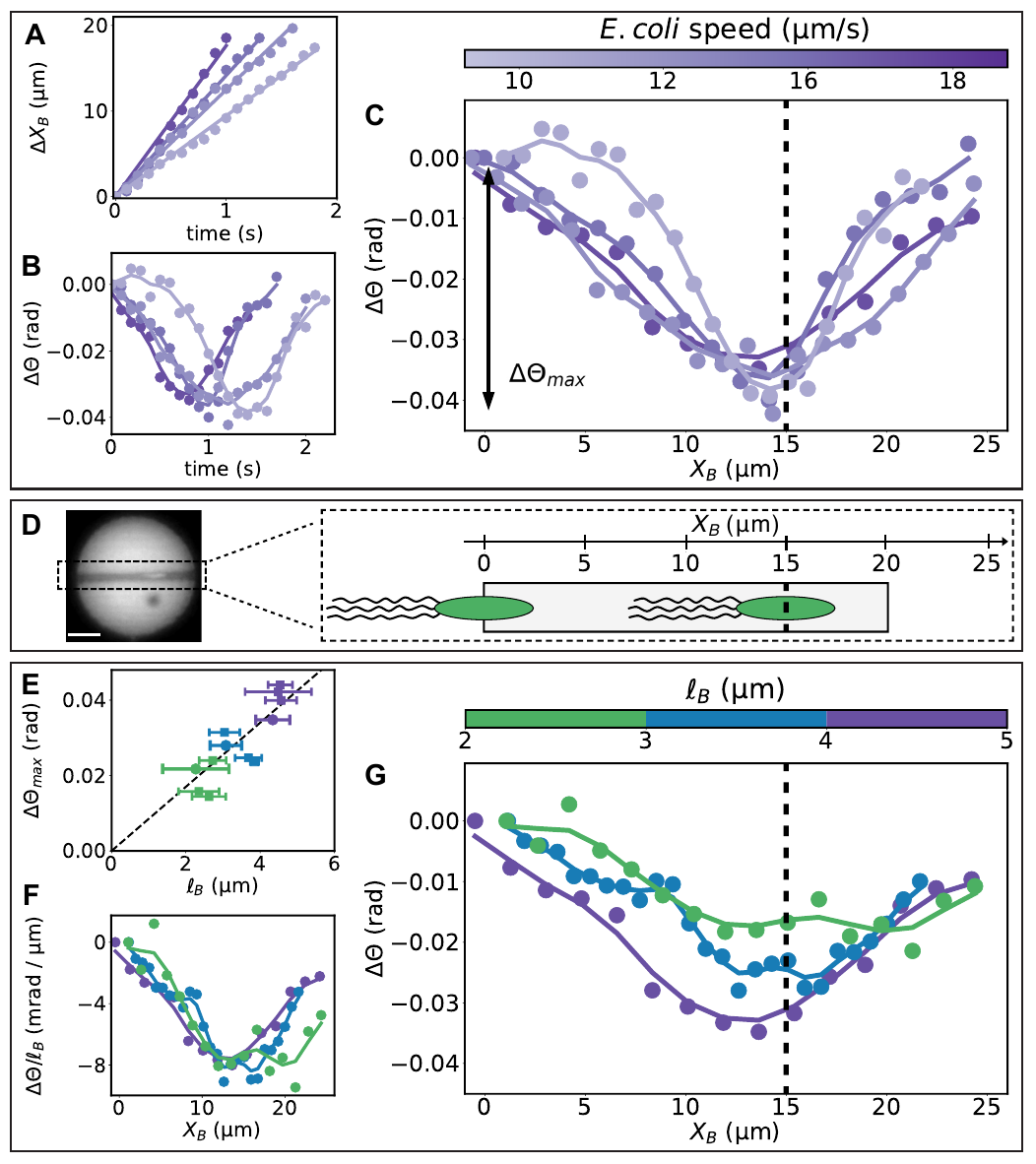} %

	\caption{\small \textbf{Dynamics of pucks with  single swimming \textit{E. coli}  crossing the channel}. (\textbf{A}) Position of swimming \textit{E. coli} ($X_B(t)$) inside the channel. Different colors indicate different bacteria, whose velocity is obtained by linear fit (solid line).  (\textbf{B}) Rotation  of the puck, [$\Delta\Theta(t)=\Theta(t)-\Theta(0)$], as a single swimming \textit{E. coli} passes through the channel. Different colors indicate different bacteria with different velocity [see: colorbar]. 
    (\textbf{C}) The curves $\Delta\Theta(t)$ from [panel (B)] collapse when represented as $\Delta\Theta(X_B)$, as prescribed by low-Reynolds-number dynamics. $\Delta\Theta(X_B)$ decreases before reversing direction, presenting a characteristic ``swoosh" shape [see main text]. The dashed line highlights the location of the minimum: $X_B=2R-\ell_B\sim$ {\SI{15}{\micro\meter}}. The depth of the swoosh is denoted $\Delta\Theta_{max}$
    (\textbf{D}) Schematic representation of \textit{E. coli} displacing in the channel. Body and flagella are at scale for a puck of \SI{10}{\micro\meter} radius. The dashed line present on (C) and position of the bacterium is also represented. Scale bar \SI{5}{\micro\meter}.
    (\textbf{E}) Depth of the swoosh, $\Delta\Theta_{max}$, as a function of the  body length of \textit{E. coli} ($\ell_B$). 
    (\textbf{F}) Collapse of the first part of the swoosh by representing ($\Delta \Theta (X_B)/ \ell_B$), as predicted by the model [see main text].
    (\textbf{G}) Rotation of the puck ($\Delta \Theta(X_B)$) for \textit{E. coli} of different sizes ($\ell_B$). The data do not collapse, in contrast to bacteria with different speeds, panel (C). For larger \textit{E. coli}, the minimum is deeper and occurs earlier. 
    In all panels, solid lines are Gaussian extrapolation of the experimental data (solid dots).
    }
	\label{fig2} 
\end{figure}

\begin{figure} 
	\centering
	\includegraphics[width=1\textwidth]{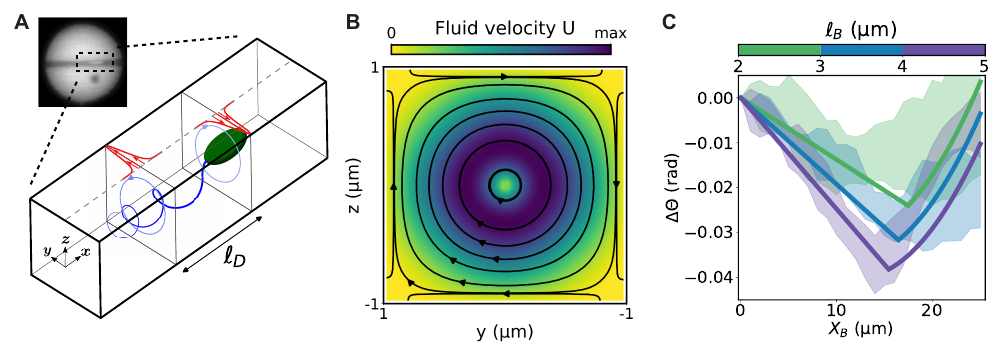} %

	\caption{\small \textbf{Hydrodynamic model  of swimming \textit{E. coli} passing through a channel.}
	   (\textbf{A}) Sketch of \textit{E. coli} swimming through a channel. Swimming \textit{E. coli} exert a torque dipole on their surrounding, stemming from the counter rotation of the cell head (clockwise when viewed from the rear) and flagella (counter-clockwise). These two torques drive a hydrodynamic flow resulting in traction forces on the top wall of the channel (red arrows). The traction fields induced by each torque  are offset by a distance $\ell_D$ and thus yield a net torque on the puck that drives the observed clockwise rotation.
        (\textbf{B}) Hydrodynamic flow field from a single, clockwise rotating rotlet near the front of the bacterium.
        (\textbf{C}) Predictions of the model for bacteria of different lengths, hence dipole separation $\ell_D$ (solid lines) and comparison with the experimental measurements (shaded zones with same color bars). The hydrodynamic model quantitatively captures the experimental observations [see main text]. 
	}
	\label{fig3} 
\end{figure}

\begin{figure} 
	\centering
	\includegraphics[width=0.8\textwidth]{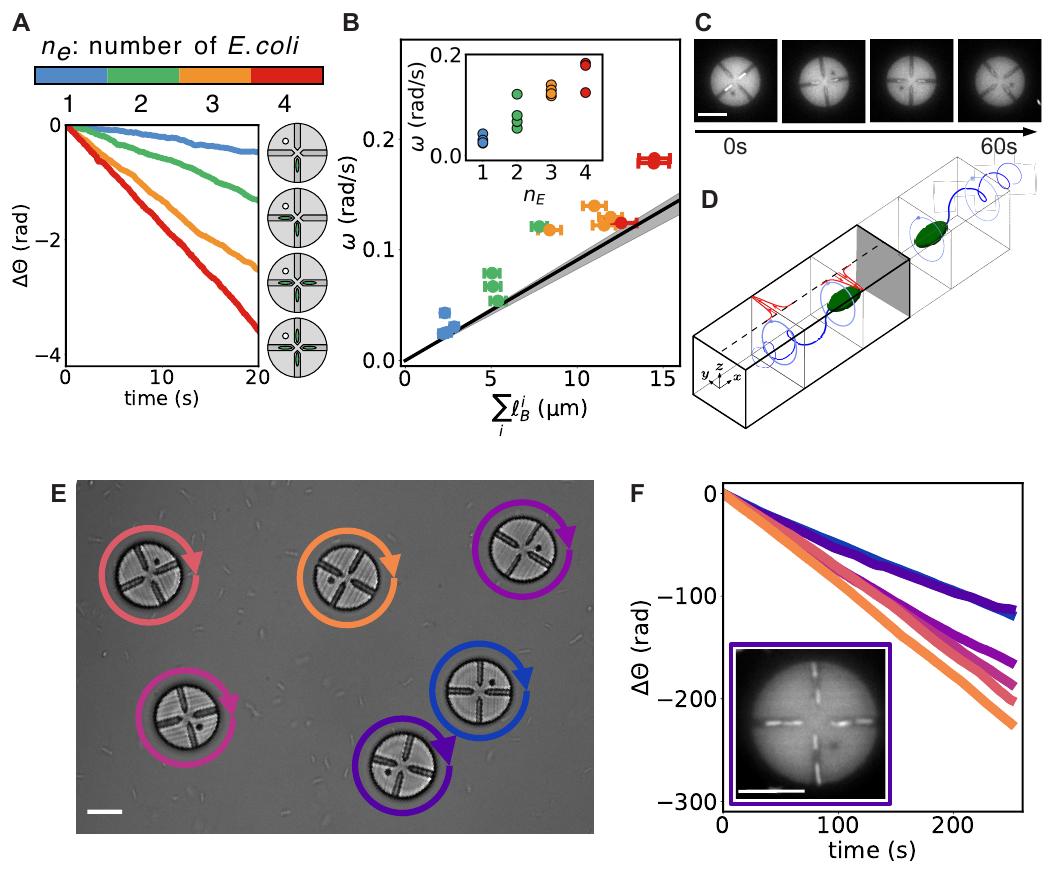} %

	\caption{\small \textbf{Dynamics of pucks with closed channels}
        (\textbf{A}) Dynamics of rotation $\Delta\Theta(t)$, for pucks containing a different number of trapped \textit{E. coli} (see schemes on the right). The  pucks all rotate clockwise, with rates increasing with number of bacteria. This highlights the cumulative effect of the entrainment mechanism. 
        (\textbf{B-inset}) Rotation rates $\omega$ for pucks with closed channels and number $n_E$ of bacteria inside. (\textbf{B}) Collapse of the data as a function of the total body length inside the channels ($\sum_i \ell_B^i$) as predicted by the model. The black line indicates the  model prediction accounting for the hydrodynamic effect of the closed channel. 
         (\textbf{C}) Timelapse of a puck with one \textit{E. coli} bacterium trapped inside each closed channel. Fluorescence images obtained using the autofluorescence of the pucks and fluorescently labeled bacteria.
        (\textbf{D}) Schematic of the hydrodynamic model for the closed channel. The no-slip condition at the channel end is implemented by placing an image system symmetrically on the other side of the wall.
        (\textbf{E}) Experiments with 6 pucks with closed channels channels immersed in a bath of motile \textit{E. coli}. All pucks spin clockwise showing a first step towards the development of chiral fluids of spinners.
        (\textbf{F}) Temporal evolution $\Delta\Theta(t)$ for the 6 pucks presented in (E). Pucks rotate persistently for minutes, at speeds up to 10 RPM. 
        (\textbf{F-inset}) Fluorescence microscopy of a puck with 4 closed channels, accommodating multiple \textit{E. coli}. It results in faster rotation rates. 
        All scale bars are \SI{10}{\micro\meter}.
        }
	\label{fig4} 
\end{figure}

\newpage 

\clearpage
\subsection*{Materials and Methods}

\subsubsection*{Stock Solutions}

For all stock solutions, chemicals were dissolved in deionized (DI) water with a resistivity of 18 M$\Omega$ cm from a Milli-Q EQ 7000 water purification system. A stock solution of 5\% w/v F108 (Sigma-Aldrich Synperonic F108 surfactant, MW 14,600) was prepared by dissolving 50g of F108 in 1L of DI water. A stock solution of 0.1 M potassium phosphate buffer (pH 8), used to prepare the motility medium, was prepared by dissolving 16.28 g of K2HPO4 (Sigma-Aldrich, MW 174.2) and 0.887 g of KH2PO4 (Sigma-Aldrich, MW 136.1) in 1L of DI water. Finally, 0.5 M of ethylenediaminetetraacetic acid (EDTA) stock solution was prepared by dissolving 186.10 g of EDTA dihydrate (Sigma-Aldrich, MW 372.2) into 1L of DI water. The pH was adjusted to 8 using NaOH pellets, to dissolve the EDTA. Stock solution of sterile Ampicillin 100mg/mL is purchased from Sigma-Aldrich. 1M L-Serine solution is made by dissolving 1.05g L-Serine powder (Sigma-Aldrich, MW = 105.09g/mol) into 10mL of DI water. 

\subsubsection*{\textit{E. coli} growth}

\textit{E. coli} (strain MG1655) are labeled with green fluorescent protein (GFP) by in house transformation using a standard electroporation protocol and a DNA plasmid containing both the GFP gene and Ampicillin resistance gene.

Cultures are grown from single colonies on agar plates containing 100ug/mL Ampicillin. Cultures are grown overnight until saturation at 33C, shaken at 200 RPM, in Tryptone Broth containing 10g/L Tryptone, and 5g/L NaCl. The saturated culture is diluted 1:100 into fresh Tryptone Broth and grown at 33C until optical density (OD) = 0.5, corresponding to mid-exponential growth phase. 1mL of cells are centrifuged at 2200 rpm for 10 min, until a pellet forms.

Cells are washed and gently resuspended in Motility Medium (MM), containing 1mM Potassium Phosphate buffer (pH 8) and 0.1mM EDTA (pH 8). This process is repeated twice more, to ensure the growth media is sufficiently removed. In the final step, the OD is remeasured and the final volume of MM is adjusted, yielding a concentrated sample of motile \textit{E. coli} at OD = 1. 

\subsubsection*{Preparation of custom 3D printed pucks}

3D printed pucks are printed using a commercial high resolution 2-Photon Polymerization (2PP) 3D printing system (NanoOne, UpNano). The pucks are designed using OnShape CAD software. Glass coverslips (No. 1.5H high precision, 170 $\pm$ \SI{5}{\micro\meter} , Zeiss) are cleaned via 10 minute sonication in 1\% Hellmanex Solution, followed by 10 minute sonication in DI water and finally plasma cleaned for 10 minutes. A 50x50 array of pucks are printed directly onto the glass coverslip ("bottom up" printing mode) using a 40x objective (NA= 1.4 WD = \SI{130}{\micro\meter}), UpBrix printing resin, and standard-fine resolution printing parameters; leftover resin is cleaned by 10 minute soak in PGMEA solution ((1-methyl-2-propyl) acetate, Sigma-Aldrich), followed by 2 minute soak in 2-Propanol (99.5\% pure, Sigma-Aldrich). 

Finally, pucks are transferred to a 50mL Falcon tube containing 5\% F-108 stock solution, and sonicated for 10 minutes to detach them from the glass coverslip. The coverslip is subsequently removed, and pucks are allowed to sediment to the bottom. Finally, 1mL is pipetted from the bottom of the tube, yielding a concentrated suspension of pucks dispersed in  5\% F-108 surfactant. This process is illustrated in Figure~\ref{figSPucks}. 

\subsubsection*{Pucks in solution}

The concentrated solution of \textit{E. coli} suspended in motility media (MM) is first diluted into MM to the desired concentration, up to a maximum of $\rho = 6 \times 10^8 $ cells/mL. L-serine stock solution (1M) is added to enhance \textit{E. coli} motility, to a final concentration of 50mM. Finally, pucks suspended in 5\% F-108 surfactant are added to a final concentration of 0.25\% F-108. The solution is confined in a 3mm x 0.3mm x 50mm rectangular glass capillary, and placed on a glass slide sealed with a wax pen; the glass capillary is previously cleaned by sonication in 1\% Hellmanex solution and washed by sonication in DI water.


\clearpage 
\bibliography{refs} 
\bibliographystyle{sciencemag}


\newpage
\section*{Acknowledgments}
We thank E. Krasnopeeva for help with the bacterial culture, motility and genetic engineering. We thank Q. Martinet for help with experimental design.
\paragraph*{Funding:}
This project has received funding from the European Research Council (ERC) under the
European Union’s Horizon 2020
research and innovation programme (VULCAN, 101086998).

\paragraph*{Author contributions:}
DG and JP conceived the experiment. DG performed and analyzed the experiment. TD and DS developed and analyzed the model. All authors wrote, reviewed and commented the manuscript.

\paragraph*{Competing interests:}
The authors declare no competing interests.

\paragraph*{Data and materials availability:}
The datasets generated and/or analyzed during the current study are openly available in the Zenodo repository, at https://doi.org/10.5281/zenodo.15236674. All data are released under the CC‑BY 4.0 license. For any further questions about data access or reuse, please contact the corresponding author.




\subsection*{Supplementary materials}
Captions for Movies S1 to S4\\
Supplementary Text\\
Figures S1 to S6\\
Table S1\\
References \textit{(7-\arabic{enumiv})}\\ 
Movies S1 to S4\\


\newpage


\renewcommand{\thefigure}{S\arabic{figure}}
\renewcommand{\thetable}{S\arabic{table}}
\renewcommand{\theequation}{S\arabic{equation}}
\renewcommand{\thepage}{S\arabic{page}}
\setcounter{figure}{0}
\setcounter{table}{0}
\setcounter{equation}{0}
\setcounter{page}{1} 


\begin{center}
\section*{Supplementary Materials for\\ \scititle}


Daniel Grober$^{1}$,Tanumoy Dhar$^{2}$,David Saintillan$^{2}$, Jeremie Palacci$^{1\ast}$\\
\small$^{1}$Institute for Science and Technology, Austria, Klosterneuburg, Austria\\
\small$^{2}$ Department of Mechanical and Aerospace Engineering, University of California San Diego, CA, USA\\
\small$^\ast$Corresponding author. Email: jeremie.palacci@ista.ac.at

\end{center}

\subsubsection*{This PDF file includes:}
Captions for Movies S1 to S4\\
Supplementary Text\\
Figures S1 to S6\\
Table S1\\

\subsubsection*{Other Supplementary Materials for this manuscript:}
Movies S1 to S4\\

\newpage


\clearpage 
\subsection*{Description of movie materials}
\noindent
\textbf{Movie S1.}
\textbf{Dynamics of a puck without a channel} Movie S1 displays a 5 minute (300s) timelapse of a puck without a channel, immersed in a bath of motile \textit{E. coli} of concentration $\rho_B = 6\times10^8$ cells/mL. The puck rotates slowly in the clockwise direction as \textit{E. coli} collide with the perimeter. Movie S1 is sped up 10x real time; scale bar \SI{20}{\micro\meter}.

\paragraph{Movie S2.}
\textbf{A single \textit{E. coli} swimming through a channel} Movie S2 displays a single \textit{E. coli} swimming through a \SI{2}{\micro\meter} x \SI{2}{\micro\meter} channel in a puck. As the \textit{E. coli} enters the channel, the puck initially rotates clockwise, and then reverses direction as the \textit{E. coli} exits. Movie S2 plays at real time (4s); scale bar \SI{20}{\micro\meter}. 

\paragraph{Movie S3.}
\textbf{Dynamics of a puck with four closed channels} Movie S3 displays a puck with 4 closed channels, each containing a single bacteria. The puck rotates persistently in the clockwise direction. Real time of the video is 86s. Movie S3 is sped up 4x; scale bar \SI{20}{\micro\meter}. 

\paragraph{Movie S4.}
\textbf{Collection of 6 pucks, each with four closed channels} Movie S4 displays the dynamics of 6 pucks, each with 4 closed channels, in a bacterial bath. The pucks rotate rapidly, up to 10RPM. Movie S4 exemplifies a route to investigate chiral fluids of spinners. Pucks are \SI{10}{\micro\meter} radii. Real time of the video is 250s. Movie S4 is sped up 4x.


\subsection*{Supplementary Text: Experiments}

\subsubsection*{Observation of puck dynamics}

The glass capillary is observed on a Nikon TI-2 Eclipse microscope equipped with a Crest Optics X-Light spinning disk system, Lumencore Celesta Light Engine laser source, and a 100x Nikon Objective (Oil immersion, NA=1.45), focused on the bottom plane of the sample. 3000 frames at 10 fps are captured for each puck using spinning disk confocal microscopy; the sample is excited with a 488nm laser and emitted light passes through a multiband filter set from AVR Optics (BrightLine filter set, Part Number: Celesta- DA/FI/TR/Cy5/Cy7-A); both the pucks and \textit{E. coli} are fluorescent at this wavelength. Images are captured using Prime 95B sCmos cameras from Teledyne Photometrics; a single camera is used for these experiments where the sample is excited using one light source. The observation is repeated for multiple pucks in the same capillary; We repeat the experiments with 4, 7, and 6 replicates for the radius 20, 10 and \SI{5}{\micro\meter} pucks, respectively.  

\subsubsection*{Differential Dynamic Microscopy}
Differential Dynamic Microscopy (DDM) \cite{Cerbino.2008, Wilson.2011} is used to quantify the speed of the \textit{E. coli} in videos used in Figure~\ref{fig1}. 
Fluorescent images of both pucks and GFP-labeled \textit{E. coli} are split into 500x500 pixel quadrants and analyzed independently, providing 4 measurements for each video; to extract the dynamics of only the \textit{E. coli}, parameters are analyzed within a narrow region of reciprocal space q ($0.5 < q < 1.5$). The analysis is repeated for videos of different pucks, and averages are taken over pucks with the same radius. For each puck (R= 5, 10 \SI{20}{\micro\meter}), the average speed of \textit{E. coli} falls within the range of 14 $\pm$ \SI{1}{\micro\meter / \second}. 

\subsubsection*{Observation of single swimmer passing through channel}

To capture the dynamics of a single swimming going through the puck with a channel, we capture multiple 3000 frames videos at 10 fps; the videos are manually segmented into 40 frame clips which capture a single swimmer entering the channel. This process was repeated for 2 pucks in the same capillary. Videos are taken using excitation at both 488nm and 640nm; both the puck and the \textit{E. coli} are fluorescent at 488nm, and only the puck is fluorescent at 640nm. A beam splitter at 525nm separates the emitted light from the sample to two separate Prime 95B sCmos cameras (Teledyne Photometrics). These images are processed later to independently track the puck and location of \textit{E. coli} in the channel. 

\subsubsection*{Image Processing - Trajectory of pucks}

To capture the angle of the pucks, we develop tracking software in Python. In each frame, we extract the center of mass of the puck and the location of the small dot by applying a Difference of Gaussian filter; the radius of the filter is adjusted to detect either the puck or the small dot. A line is drawn between these two points, and the angle of this line with respect to the horizontal gives the orientation of the puck, as seen in the inset in Figure~\ref{fig1}B. The angular speed of the pucks is given by a linear regression between this orientation and time. To quantify the rotational diffusion of the pucks, we rotate the trajectory into this rotating frame of reference and calculate the Mean Squared Angular Displacement (MSAD); the rotational diffusion is given by a linear regression between MSAD and time, up to 1.5s. 

\subsubsection*{Image Processing - \textit{E. coli} swimming through the channel}

For the experiments where a single \textit{E. coli} swims through the channel, we first track the puck using the Difference of Gaussian filter described previously. We track the orientation of the puck using only light greater than 525nm (red channel), to decouple locating of the center of mass of the puck from any light emitted from the \textit{E. coli} as it swims through the channel. Next, we rotate the image into the frame of reference of the puck, such that the channel is placed horizontally. The \textit{E. coli} is tracked using emitted light less than 525nm (green channel). We convolve the image with a Difference of Gaussian filter of size equal to the width of the \textit{E. coli} and apply a binary mask to the image, revealing only the contents of the channel. Finally, the image is binarized; the length of the major axis of the binary region is taken as the size of the cell ($\ell_B$); the orientation of the major axis with respect to the horizontal is taken as the angle of the \textit{E. coli} inside the channel ($\Phi_e$). For images where the \textit{E. coli} cannot be effectively identified using the image processing techniques described above, the position cell body is located by clicking on the image. Finally, we track the center of the cell body from frame to frame, and extract a speed for each \textit{E. coli}, as displayed in Figure~\ref{fig2}A.

In Figure~\ref{figSCorr}, we display data for 12 individual instances where a single \textit{E. coli} runs through the puck. For each instance, we track 4 statistics to describe the event: $\Delta\Theta_{max}$ vs, the maximum change in angle of the puck over the timeframe where the \textit{E. coli} is inside the channel, $U_s$, the velocity of the \textit{E. coli}, $\langle\Phi_e\rangle$, the average angle of the \textit{E. coli} cell body with respect to the channel, and $\ell_B$, the length of the \textit{E. coli} cell body. In Figure~\ref{figSCorr}A we display a correlation matrix between these statistics; we find that $\Delta\Theta_{max}$ and $\ell_B$ are strongly correlation. In Figure~\ref{figSCorr}B and Figure~\ref{figSCorr}C, we plot $\Delta\Theta_{max}$ vs $U_s$ and $\langle\Phi_e\rangle$, respectively. These variables do not display a strong correlation. Finally, in Figure~\ref{figSCorr}D we plot $U_s$ and $\ell_B$; these variables do not display a strong correlation.  

\subsubsection*{Rotational diffusion of puck in thermal bath}

We measure the rotational mobility of the \SI{10}{\micro\meter} puck by tracking the orientation of the puck in a thermal bath ($\rho_B$ = 0).  The experimental conditions are otherwise exactly identical to experiments including the bacterial bath; the pucks are suspended in Motility Media, 0.25\% F-108 surfactant and 50mM L-serine, and sealed in a glass capillary. Because the rotation is very slow owing for an object of \SI{10}{\micro\meter} radius, acquisition is performed over the course of a few hours. The puck is observed for 10,000 seconds at a frame rate 0.1 fps (1000 frames) using confocal microscopy. Figure~\ref{figSPuckThermal}A displays the angle of the puck as a function of time; this data is used to calculate an average Mean Squared Angular Displacement ($\langle \Delta \Theta ^2 \rangle$) at timesteps up to $\Delta t$ = 1000s, as displayed in Figure~\ref{figSPuckThermal}B. A rotational diffusion coefficient ($D_\Theta$) for the puck is determined by a linear fit between $\langle \Delta \Theta ^2 \rangle$ and $\Delta t$. {We find  $D_\Theta$ = $6\times10^{-5} \pm 1\times10^{-5}$ rad$^2$/s. The mobility of the puck is determined using the Stokes-Einstein relation: }

\begin{equation*}
    D_\Theta = M_\Theta k_BT,
\end{equation*}

\noindent with $M_\Theta$ the rotational mobility, $k_B$ the Boltzmann constant and $T$ the temperature. 

\subsubsection*{Observation of puck with partial channel}
In Figure~\ref{fig4}, we investigate the dynamics of pucks with 4 closed channels. These pucks are 3D printed as described previously, and dispersed in solution containing Motility Medium, 0.25\% F-08, and 50mM L-serine, and motile GFP-labeled \textit{E. coli} (identical to previous experiments). The concentration of \textit{E. coli} is adjusted to $1\times 10^8$ cells/mL. We capture movies at 20fps using the same 100x Nikon Objective and Nikon spinning disk confocal microscope described in previous section ``Observation of puck dynamics". Cell body lengths are measured directly from images, using the Nikon Elements software.

\newpage
\subsection*{Supplementary Text: Simulations}

We solve for the flow and traction fields induced by a point torque $\boldsymbol{\Gamma}$ inside an infinite channel using the boundary element method. By linearity of the Stokes equations (\ref{eq:stokes}), the velocity $\mathbf{U}$ in the fluid can be decomposed as a sum of two contributions: the flow $\mathbf{U}^t$ due to a point torque in free space in the absence of confinement, and a correction $\mathbf{U}^c$ calculated to satisfy the no-slip condition on the walls,
\begin{equation}
\mathbf{U}(\mathbf{r})= \mathbf{U}^t (\mathbf{r})+ \mathbf{U}^c (\mathbf{r}). 
\end{equation}
The first contribution is simply given by a rotlet flow,
\begin{equation}
\mathbf{U}^t(\mathbf{r})= \frac{1}{8\pi\mu} \frac{\boldsymbol{\Gamma}\times \mathbf{r}}{|\mathbf{r}|^3},
\end{equation}
where the position vector $\mathbf{r}$ points from the location of the singularity. The correction $\mathbf{U}^c(\mathbf{r})$ satisfies the homogeneous Stokes equations
$ \mu \nabla^2 \mathbf{U}^c-\nabla P^c  = \mathbf{0}$ and   $\nabla\cdot\mathbf{U}^c=0$, 
subject to the Neumann boundary condition $\mathbf{U}^c(\mathbf{r}) = -\mathbf{U}^t(\mathbf{r})$ for points $\mathbf{r}$ on the channel walls $S$, which ensures that the total flow field $\mathbf{U}(\mathbf{r})$ satisfies the no-slip condition. We solve for the correction $\mathbf{U}^c$ numerically using a single-layer boundary integral representation \cite{pozrikidis1992boundary},
\begin{equation}
\mathbf{U}^c (\mathbf{r}) =\frac{1}{8\pi\mu} \int_S \mathbf{G}(\mathbf{r}-\mathbf{r}_0)\cdot \mathbf{q}(\mathbf{r}_0)\,\mathrm{d}S(\mathbf{r}_0),  \label{eq:BIeq}
\end{equation}
where $\mathbf{G}(\mathbf{r})=(\mathbf{I}+\hat{\mathbf{r}}\hat{\mathbf{r}})/|\mathbf{r}|$ is the Oseen tensor, and $\mathbf{q}(\mathbf{r})$ is the unknown single-layer density defined on the domain boundaries. Evaluating Eq.~(\ref{eq:BIeq}) on the channel walls and applying the boundary condition on $\mathbf{U}^c$ yields an integral equation for the unknown single-layer density $\mathbf{q}$, which we solve numerically after discretizing the channel walls with rectangular elements. The domain of integration is truncated in the $\pm x$ direction at a distance chosen large enough to ensure convergence of the flow field, and $N=1600$ elements are used in the calculations. After obtaining $\mathbf{q}(\mathbf{r})$ on the walls, we reconstruct the flow field inside the channel using Eq.~(\ref{eq:BIeq}), and differentiate it numerically to obtain the surface traction $\mathbf{t}(\mathbf{r})$ entering Eq.~(\ref{eq:traction}) for the torque on the puck. 

{In the case of a closed channel [Fig.~\ref{fig4}], the flow solution above can be modified to satisfy the no-slip condition at the channel end as well. Assuming the channel is closed by a flat wall, the flow solution obtained above is easily modified using the method of images, by placing a counter-rotating image torque $-\boldsymbol{\Gamma}$ on the other side of the wall \cite{blakechwang}. The effect of the wall is to reduce the magnitude of the traction induced by the rotlet, with a modified value of $\Lambda$ in Eq.~(\ref{eq:gamma1}) that now depends on the distance $d$ from the rotlet to the channel end, and that we denote by $\Lambda^d$. }

When the \textit{E. coli} is inside the channel, the net torque on the puck resulting from the two rotlets is modified as
\begin{equation}\label{angv2_corr}
\Gamma = -\Lambda^{d} \left(\frac{x_1-x_C}{W}\right) \Gamma_M  +\Lambda^{d+\ell_D} \left(\frac{x_2-x_C}{W}\right) \Gamma_M,
\end{equation}
where $d$ is the distance from the wall to the first rotlet, and $d+\ell_D$ is the distance to the second rotlet. Also, $x_1-x_C=d+d_C$, where $d_C$ is the offset from the channel end to the center of the puck, which is provided by the experimental design and is on the order of $1\,\mu m$. We find that $\Lambda^d$ is a rapidly decaying function of $d$, and given the  value of $\ell_D$, $\Lambda^{d+\ell_D}\approx \Lambda$ in all cases. Because the effect of the wall is to decrease the magnitude of the traction field induced by the first rotlet, it results in an enhancement of the net torque on the puck. The factor that captures the enhancement of the torque with respect to the value in open channels is given by the ratio of Eq.~(\ref{angv2_corr}) and Eq.~(\ref{eq:nettorque}) and simplifies to
\begin{equation}
\chi=\frac{\Gamma}{- \Lambda \left(\frac{\ell_{D}}{W} \right)\Gamma_{M}} = \bigg[1 +  \frac{(d + d_{C})}{\ell_{D}}\left(1 - \frac{\Lambda^{d}}{\Lambda}\right)\bigg].  \label{eq:chi}
\end{equation}
The dependence of this ratio on the distance $d$ is plotted in Fig.~\ref{fig_SI_t_vs_d}, and shows an enhancement of $\sim 0 - 15\%$. The precise value of $d$ is unknown, but  we can estimate it to be on the order of $0.5\,\mu m$ based on the channel geometry if we approximate the \textit{E. coli} body as a cylinder of radius $0.5\,\mu m$. For this value of $d$, our model predicts a torque enhancement of $\sim 10\%$.  

\newpage

\begin{figure} 
	\centering
	\includegraphics[width=0.9\textwidth]{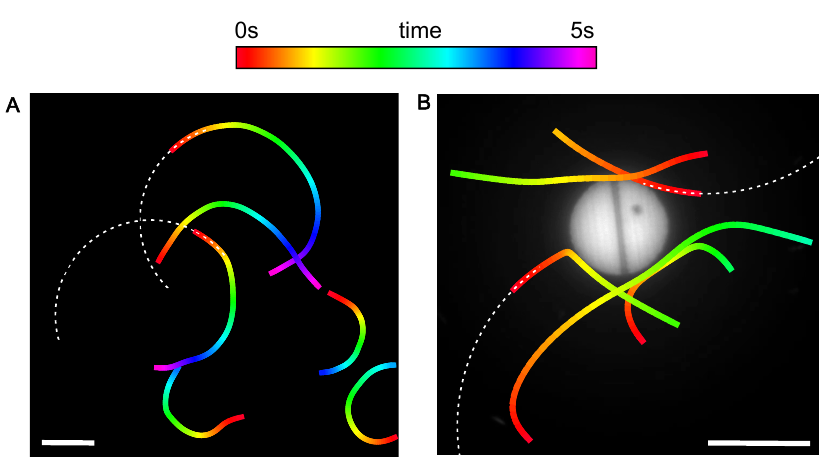} 

	\caption{\small \textbf{Trajectories of \textit{E. coli} swimming near a no-slip boundary condition} 
    \textbf{(A)} \textit{E. coli} swim in clockwise circular trajectories when swimming above a glass interface. Each colored line represents the trajectory of a single \textit{E. coli}; time is indicated by the colorbar. The white dashed curves represents a circle of radius \SI{40}{\micro\meter}. Scale bar \SI{20}{\micro\meter}.
    \textbf{(B)} Each colored line represents the trajectory of a single \textit{E. coli} which collides with the exterior of the puck. The \textit{E. coli} swim in clockwise circular trajectories and scatter off the exterior of the puck. Scale bar \SI{20}{\micro\meter}.
     	}
		
	\label{figSTrajectories} 
\end{figure}

\begin{figure} 
	\centering
	\includegraphics[width=1\textwidth]{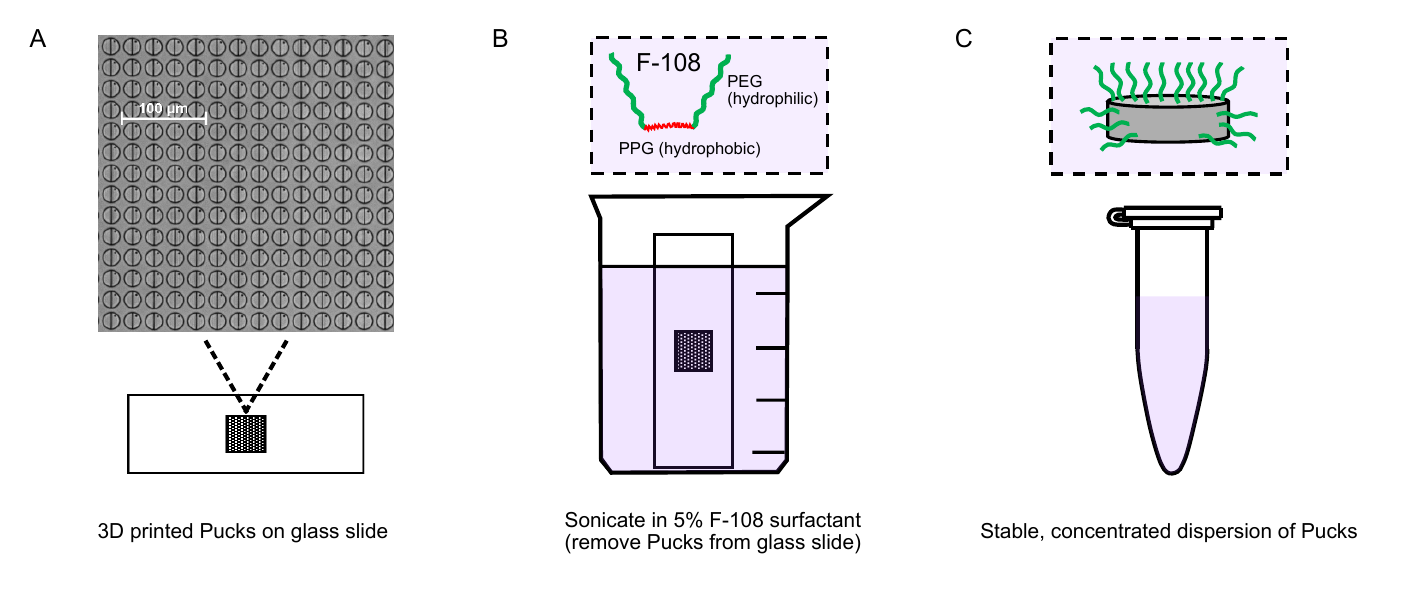} 
	\caption{\small \textbf{3D printing and dispersion of pucks}
    \textbf{(A)} Pucks are intially 3D printed on a glass coverslip, using 2-Photon-Polymerization technique, in 50x50 arrays. 
    \textbf{(B)} To remove the pucks from the glass, the coverslip is placed in a 50mL Falcon tube containing 5\% F-108 surfactant, and sonicated. F-108 is a tri-block copolymer, containing two hydrophilic sections (PEG) and a hydrophobic section in the middle (PPG).
     \textbf{(C)} The glass coverslip is subsequently removed, and pucks are allowed to sediment to the bottom. Finally, a concentrated dispersion of pucks is obtained by pipetting the bottom 1mL of solution into an eppendorf tube. 
		}
		
	\label{figSPucks} 
\end{figure}

\begin{figure} 
	\centering
	\includegraphics[width=0.5\textwidth]{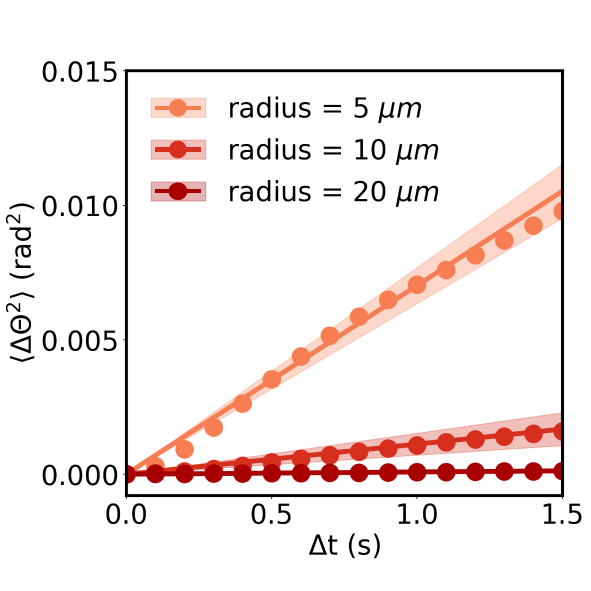} 
	\caption{\small \textbf{Mean Squared Angular Displacement of pucks without Channel}
    Mean Squared Angular Displacement ($\langle \Delta \Theta^2 \rangle$) is calculated by first moving the angular trajectory of the puck into a rotating reference frame, equal to its average angular velocity; $\langle \Delta \Theta^2 \rangle$ is calculated by taking the average change in angle of the puck at time steps ($\Delta t$), ranging from 0.1 to 1.5 s. The Rotational Diffusion of the puck is quantified via a linear fit between  $\langle \Delta \Theta^2 \rangle$ and $\Delta t$. In the above figure, dots represent the trajectory of a single puck, while solid lines and shaded regions represent the average and standard deviation over all pucks of that size. We capture 4,7, and 6 pucks of sizes R=5,10 and \SI{20}{\micro\meter}, respectively. 
		}
	\label{figSMSAD} 
\end{figure}

\begin{figure} 
	\centering
	\includegraphics[width=0.75\textwidth]{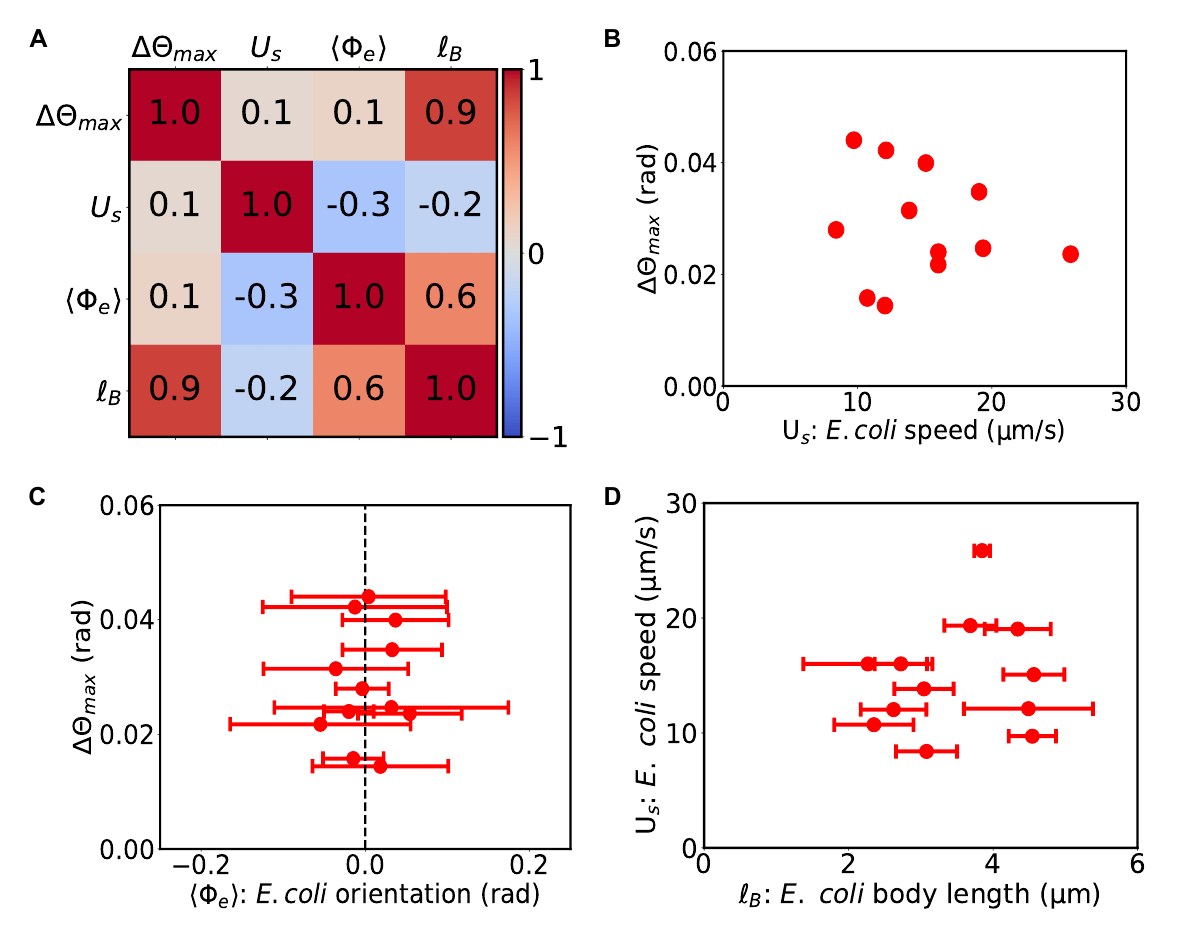} 
	\caption{\small \textbf{Comparison of correlation of the maximal rotation  $\Delta \Theta_{max}$ of the pucks [see main text] with statistical or dynamical properties of \textit{E. coli}}. In order to understand the mechanism of rotation of the puck, we identify parameters that intuitively could induce the rotation of the puck and investigate their correlation. Following past work that showed the rotation of microgears in bacterial baths due to collisions, we notably investigate the effect of the speed of the \textit{E. coli} and their average orientation in the channel. As visible from (A), the correlation is weak, highlighting that our mechanism for rotating the pucks is different than previously reported. Remarkably, it correlates well with the length of the bacterium body, as explained by our model of torque dipoles [see main text]. \textbf{(A)} Each time an E. coli swims through the channel, the rotation of the puck is quantified using the maximum change in angle of the puck ($\Delta \Theta_{max}$) during interval where \textit{E. coli} is inside the channel. The motion of the \textit{E. coli} is quantified using 3 statistics: its speed ($U_s$), its average angle with respect to the channel ($\langle \Phi_E \rangle$), and the length of the cell body ($\ell_B$). The correlation matrix displays the Pearson correlation coefficient between such variables.
    \textbf{(B)} Maximum change in angle of the puck ($\Delta \Theta_{max}$) and \textit{E. coli} speed ($U_s$); each dot represents a single \textit{E. coli} swimming through the channel. No significant correlation is found between $\Delta \Theta_{max}$ and $U_s$. 
     \textbf{(C)} Maximum change in angle of the puck ($\Delta \Theta_{max}$) and average angle of the \textit{E. coli} with respect to the channel ($\langle \Phi_E \rangle$). No significant correlation is found between $\Delta \Theta_{max}$ and $\langle \Phi_E \rangle$. 
     \textbf{(D)} \textit{E. coli} speed ($U_s$) and length of the cell body ($\ell_B$). No significant correlation is found between $U_s$ and $\ell_B$. }
		
	\label{figSCorr} 
\end{figure}
\begin{figure} 
	\centering
	\includegraphics[width=1\textwidth]{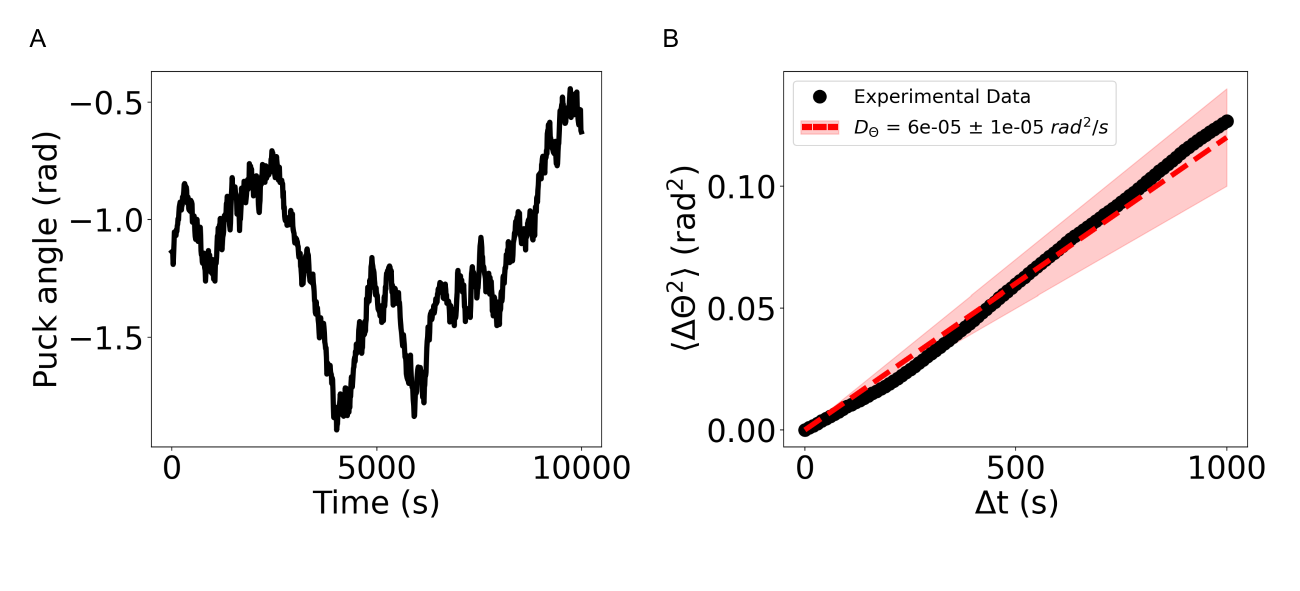} 
	\caption{\small \textbf{Rotational dynamics of a puck in thermal bath}
    \textbf{(A)} The orientation of a single puck (R=\SI{10}{\micro\meter}) in a thermal bath is tracked for 10,000 seconds, using a frame rate of 0.1fps. 
    \textbf{(B)} Average Mean Squared Angular Displacement ($\langle \Delta\Theta^2\rangle$) as a function of timestep ($\Delta t$) is calculated using the trajectory shown in \textbf{(A)}. A rotational diffusion coefficient for the puck ($D_\Theta$) is extracted via linear fit between $\langle \Delta\Theta^2\rangle =2D_\Theta \Delta t$, and visualized with the red dashed line; a confidence interval representing estimated standard measurement error is displayed in red shaded region. We determine the rotational diffusion coefficient to be $D_\Theta = (6 \pm 1)\times 10^{-5}$ rad$^2$/s
     	}
		
	\label{figSPuckThermal} 
\end{figure}

\begin{figure} 
	\centering
	\includegraphics[width=0.65\textwidth]{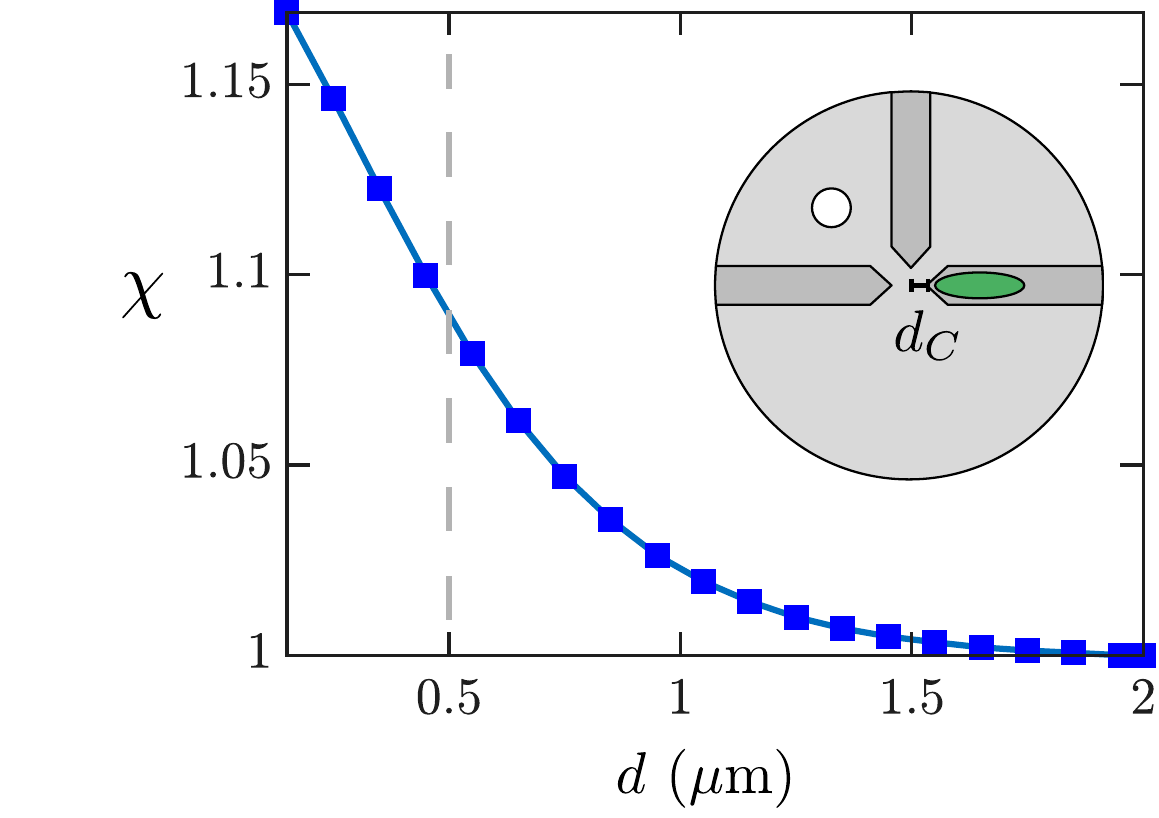} %

	\caption{\small \textbf{Enhancement $\chi$ of the torque in a closed channel as a function of the distance $d$ from the wall.} In a closed channel, the no-slip condition at the channel end results in a decrease in the traction induced by the first rotlet, and thus causes an increase in the net torque on the puck by a factor $\chi$ estimated in Eq.~(\ref{eq:chi}). The distance $d$ from the first rotlet to the wall is estimated to be $\sim 0.5\,\mu$m in the experiment (vertical dashed line). }
	\label{fig_SI_t_vs_d} 
\end{figure}

\begin{table} 
	\centering
    \caption{\textbf{Choice of parameters for simulations}
		The following table includes values used to fit our hydrodynamic simulations to the experimental work, notably in Fig.~3C and Fig.~4B. \\
		}
	\label{STable_params} 

    \begin{tabular}{
    |p{4cm}||p{3cm}|p{3cm}|p{4cm}|  }
     \hline
     \multicolumn{4}{|c|}{\textbf{Choice of parameters}} \\
     \hline
     Variable & References & Range & Current work\\
     \hline
     Motor torque $ {\Gamma}_{M}$   & \cite{das2018computing}
     \cite{berry1997absence}
     \cite{fahrner2003bacterial}
     &400-4500 pN-nm&   1600 pN-nm \\
     \hline
    Stokesian mobility of the puck $ {M_{\Theta}}$ &   Supplemental material, [Fig.~\ref{figSPuckThermal}]  & [1.2,1.7]$\times10^{-5} \newline$  (sec-pN-nm$)^{-1}$  &$2\times10^{-5}$ (sec-pN-nm$)^{-1}$\\
    \hline
     Swim speed of the bacterium $U^{s}$ & [Fig.~\ref{figSCorr}D],\newline
    \cite{Grober.NatPhys.2023}& 5-\SI{25}{\micro\meter / \second} &  \SI{15}{\micro\meter / \second}\\
    \hline
     Puck radius $ {R}$   &Main Text \newline [Fig.~\ref{fig1}D] & 5-\SI{20}{\micro\meter} &  \SI{10}{\micro\meter} \\
     \hline
     Channel height $H$  &  Main Text & \SI{2}{\micro\meter} &\SI{2}{\micro\meter}\\
     Channel width $W$& [Fig.~\ref{fig1}D]  & \SI{2}{\micro\meter}   & \SI{2}{\micro\meter} \\
     \hline
    Rotlet dipole size $\ell_{D}$ & Main text, \newline [Fig.~\ref{figSCorr}D]  & $\alpha \times \text{cell body}, \newline \alpha = 1.5$  & 1.5 $\ell_{B}$\\
     \hline
    \end{tabular}
    
\end{table}



\end{document}